\documentclass[
    ,final            
  ]
  {aipproc}

\layoutstyle{6x9}

\begin{document}

\title{Global Analysis of Helicity PDFs: \\
past - present - future\footnote{talk presented by M.\ Stratmann}}

\classification{}
\keywords{}

\author{D.\ de Florian}{
  address={Instituto de F\'{\i}sica de Buenos Aires and 
Departamento de F\'{\i}sica, Facultad de Ciencias Exactas y Naturales,
Universidad de Buenos Aires, Buenos Aires, Argentina}
}

\author{R.\ Sassot}{
  address={Instituto de F\'{\i}sica de Buenos Aires and 
Departamento de F\'{\i}sica, Facultad de Ciencias Exactas y Naturales,
Universidad de Buenos Aires, Buenos Aires, Argentina}
}

\author{M.\ Stratmann}{
  address={Physics Department, Brookhaven National Laboratory, Upton, NY 11973, USA}
}

\author{W.\ Vogelsang}{
  address={Institut\ f\"{u}r Theoretische\ Physik, Universit\"{a}t T\"{u}bingen, 72076 T\"{u}bingen, Germany}
}

\begin{abstract}
We discuss the current status of the DSSV global analysis of
helicity-dependent parton densities. A comparison with recent semi-inclusive DIS
data from COMPASS is presented, and constraints on
the polarized strangeness density are examined in some detail.
\end{abstract}

\maketitle

\section{Introduction: DSSV global analysis}
Helicity-dependent parton densities (PDFs) tell us precisely 
how much quarks and gluons with a given momentum fraction $x$ tend to have their 
spins aligned with the spin direction of a nucleon in a helicity eigenstate. 
Their knowledge is essential in the quest to answer one of the most basic 
questions in hadronic physics, namely how the spin of a nucleon
is composed of the spins and orbital angular momenta of its constituents.

More than a dozen experiments have measured with increasing precision 
various observables sensitive to different combinations of quark and gluon polarizations 
in the nucleon.
The experimental progress was matched by advancements in corresponding theoretical 
higher order calculations in the framework of pQCD and 
phenomenological analyses of available data \cite{ref:dssv,ref:otherfits}.
The most comprehensive global fits \cite{ref:dssv} include data taken in spin-dependent
DIS, semi-inclusive DIS (SIDIS) with identified pions and kaons, and 
proton-proton collisions. They allow for extracting sets of helicity PDFs 
consistently at next-to-leading order (NLO) accuracy along with estimates of their uncertainties.
Contributions from the orbital angular momenta of quarks and gluons completely decouple from
such type of experimental probes and need to be quantified by other means. 
One important asset in the DSSV global analysis framework is the use of a numerically fast Mellin moment technique \cite{Stratmann:2001pb,ref:dssv}
which allows one to incorporate complicated NLO
expressions for $pp$ processes \cite{ref:polnlo} without any approximations.

Unlike unpolarized PDF fits, where a separation of different quark flavors is obtained 
from inclusive DIS data taken with neutrino beams, differences in polarized quark and antiquark densities are at present determined exclusively from SIDIS data and hence require knowledge of fragmentation functions (FFs). 
Reliable sets of FFs at NLO accuracy have been extracted in global fits to 
inclusive hadron yields in $e^+e^-$, $ep$, and $pp$ collisions \cite{ref:dss}. 
Even though pion FFs are rather well constrained by data, corresponding kaon FFs suffer 
from larger uncertainties which complicate current extractions of $\Delta s(x)$.

\section{Recent DIS and SIDIS data}
%
\begin{figure}
  \includegraphics[width=.4\textwidth]{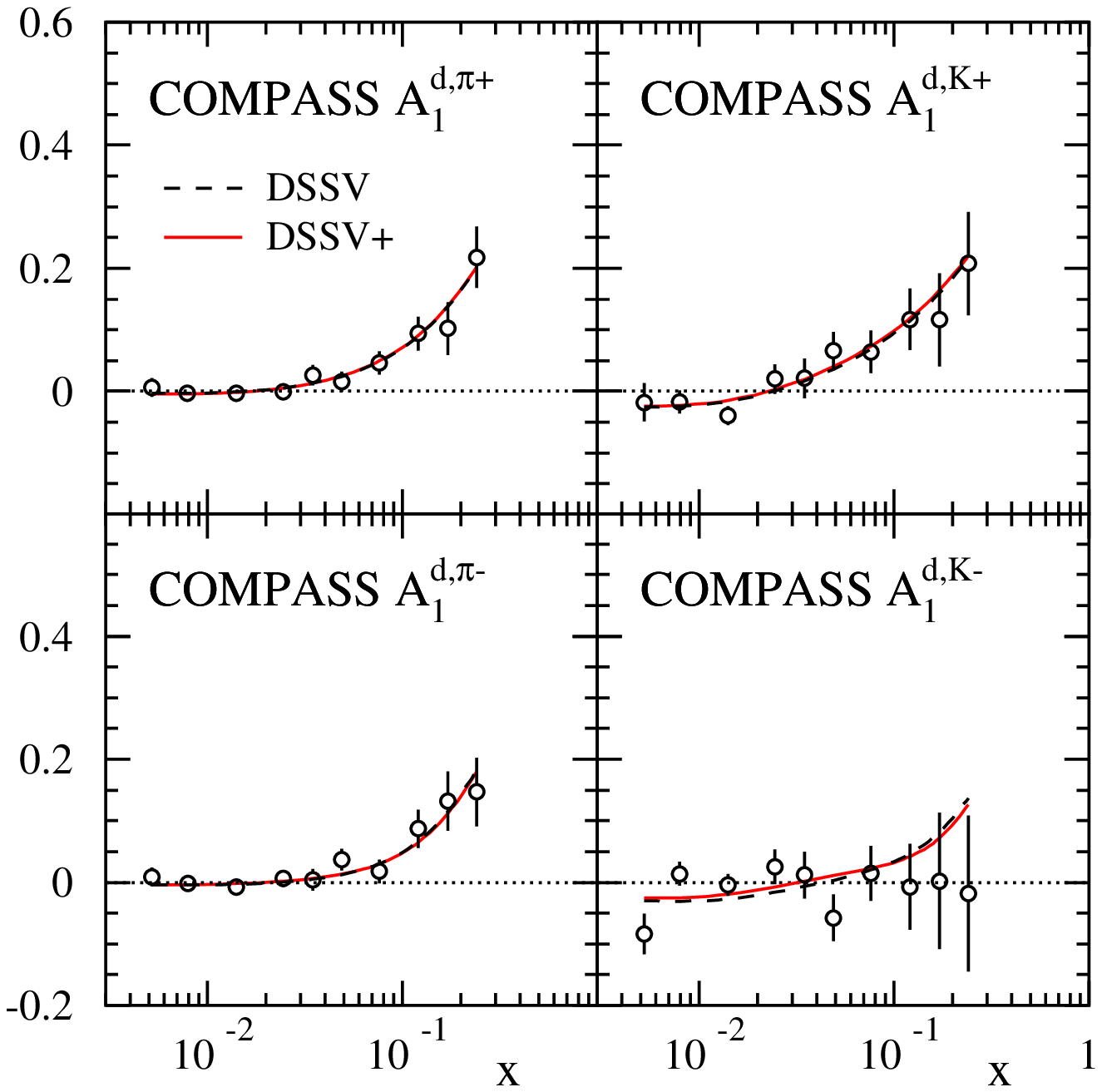}
  \includegraphics[width=.4\textwidth]{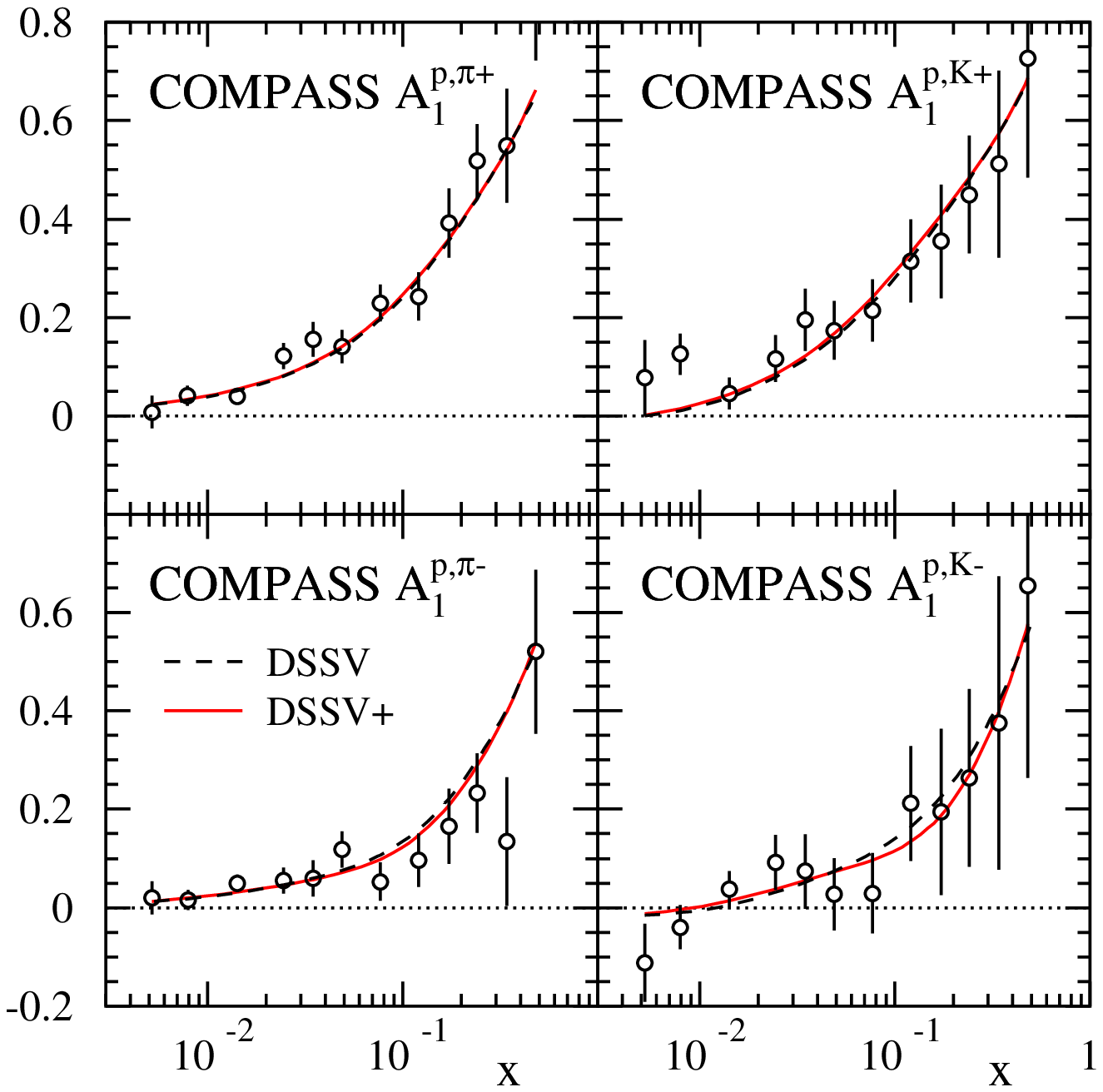}
\caption{\label{fig:newsidis} COMPASS results~\cite{Alekseev:2009ci,Alekseev:2010ub} for SIDIS spin asymmetries on a deuteron (left) and proton target (right) compared to DSSV and DSSV+ fits (see text).}  
\end{figure}
Recently, the COMPASS collaboration has published
new DIS~\cite{Alekseev:2010hc} and SIDIS~\cite{Alekseev:2009ci,Alekseev:2010ub} data. 
The latter extend the coverage in $x$ down to about $x\simeq 5\times10^{-3}$, almost
an order of magnitude lower than the kinematic reach of the HERMES data 
used in the DSSV global analysis of 2008~\cite{ref:dssv}.
For the first time, the new results comprise measurements of identified pions and 
kaons taken with a longitudinally polarized proton target.
Clearly, these data can have a significant impact on fits of helicity PDFs 
and estimates of their uncertainties.
In particular, the new kaon data will serve as an important check of the validity of the strangeness density obtained in the DSSV analysis, which instead of favoring a negative polarization as in most fits based exclusively on DIS data, prefers a vanishing or perhaps even slightly positive $\Delta s$ in the measured range of $x$, see below.

The new data for the inclusive spin asymmetry 
$A_1^p$ appear to be well described by the original DSSV set
of helicity PDFs yielding a $\chi^2/\mathrm{d.o.f.}\approx 1$.
Figure~\ref{fig:newsidis} shows a detailed comparison between the new SIDIS spin
asymmetries from COMPASS~\cite{Alekseev:2009ci,Alekseev:2010ub} and
the original DSSV fit (dashed lines).
Also shown is the result of a re-analysis at NLO accuracy (denoted as ``DSSV+'')
based on the updated data set.
The differences between the DSSV and the DSSV+ fits are hard to notice 
for both identified pions and kaons. 
The total $\chi^2$ of the fit drops only by a few units upon refitting, 
which is not really a significant improvement for a PDF analysis
in view of non-Gaussian theoretical uncertainties. The change in $\chi^2$ is also
well within the maximum $\Delta \chi^2/\chi^2=2\%$
tolerated as a faithful, albeit conservative estimate of PDF uncertainties
within the DSSV global analysis \cite{ref:dssv}.

At first sight it may seem that the new SIDIS data have only very little impact on the fit.
This is not the case if one studies individual $\chi^2$ profiles in more detail.
Compared to the original DSSV fit we find a trend towards smaller net polarization 
for $\Delta \bar{u}$ and $\Delta \bar{d}$ in the range $0.001\le x \le 1$.
In addition, one finds a significant reduction in the uncertainties, as determined by the width of the $\chi^2$ profiles at a given $\Delta \chi^2$.
There is, however, some mild tension with older SIDIS sets, but this is well within the tolerance of the fit and most likely caused by the different $x$ ranges covered by the different data sets. 

\subsection{Constraining the strangeness helicity density}
\begin{figure}
  \includegraphics[width=.6\textwidth]{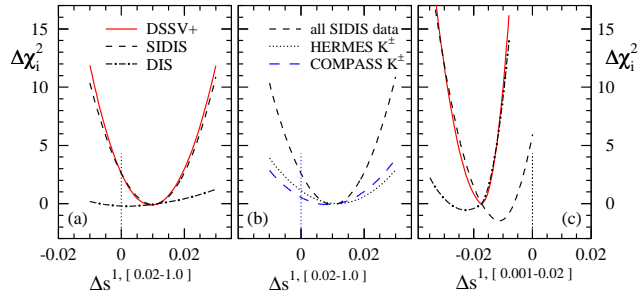}
  \vspace*{-0.5cm}
  \caption{\label{fig:profiles-s} (a), (c): $\chi^2$ profiles for the truncated first moment
  of $\Delta s$ in two different $x$ intervals. (b): impact of kaon data from COMPASS and
  HERMES in the range $0.02\le x \le 1$.}  
\end{figure}

A much debated feature of the strangeness helicity PDF obtained in the DSSV fit is 
its unexpected small value at medium-to-large $x$ which, 
when combined with a node at intermediate $x$, 
still allows for acquiring a significant negative first moment at small $x$,
in accordance with expectations from SU(3) symmetry and fits to DIS data only.
To investigate the possibility of a node in $\Delta s(x)$ further, we present
in Fig.~\ref{fig:profiles-s}
the $\chi^2$ profiles for two different intervals in $x$: (a) 
$0.02\le x \le 1$ and (c) $0.001\le x \le 0.02$. The middle panel (b) demonstrates
the impact and consistency of kaon data from HERMES and COMPASS in constraining
$\Delta s(x)$ in the region $0.02\le x \le 1$.

The profiles in Fig.~\ref{fig:profiles-s} clearly show that the result for $\Delta s$ for $0.001\le x \le 0.02$
is a compromise between DIS and SIDIS data, the latter favoring 
less negative values. For $0.02\le x \le 1$ everything is determined by SIDIS data, and all 
sets consistently ask for a small, slightly positive strange quark polarization. 
There is no hint of a tension with DIS data here as they do not provide a useful constraint at 
medium-to-large $x$.
We note that at low $x$, most SIDIS sets in the original DSSV fit 
give indifferent results. The new COMPASS data, which extend towards the smallest $x$ values so far, actually show some preference for a slightly negative value for $\Delta s$. 
We also notice that in the range $x>0.001$ the hyperon decay constants, the so-called
$F$ and $D$ values, do not play a significant role in constraining $\Delta s(x)$.
To quantify possible SU(3) breaking effects one needs to probe $\Delta s(x)$ at smaller values 
of $x$, for instance in SIDIS at a future EIC \cite{Boer:2011fh}.

Clearly, all current extractions of $\Delta s$ from SIDIS data suffer from a significant dependence on kaon FFs, see, e.g., Ref.~\cite{Alekseev:2009ci,Alekseev:2010ub}, and better determinations of $D^K(z)$ are highly desirable. 
Contrary to other fits of FFs \cite{:2008afa},
only the DSS sets \cite{ref:dss} provide a satisfactory
description of pion and kaon multiplicities in the same kinematic range where we have
polarized SIDIS data. 

\section{Outlook}
%
Existing experiments, like PHENIX and STAR at RHIC, will continue to add data in the
next couple of years. Preliminary single-inclusive jet data from STAR 
presented at this conference exhibit a non-zero double-spin asymmetry $A_{\mathrm{LL}}$
in the covered range of transverse momenta $p_T$ \cite{ref:star09}. 
Measurements of $A_{\mathrm{LL}}$ for di-jet correlations \cite{ref:dijet}
should help to improve the current constraints on $\Delta g(x)$ and extend them towards somewhat smaller values of $x$. As soon as these data sets are finalized they will be
incorporated in our global analysis framework, and their implications for $\Delta g(x)$ 
will be studied in detail.

Parity-violating, single-spin asymmetries for $W$ boson production from RHIC
should reach a level where they help to constrain $\Delta u$, $\Delta \bar{u}$,
$\Delta d$, and $\Delta \bar{d}$ at moderately 
large $x$, $0.07\le x \le 0.4$ at scales $Q\simeq M_W$,
much larger than typically probed in SIDIS \cite{deFlorian:2010aa}. 
The strangeness polarization is, however, very hard to access in polarized $pp$ collisions.
In the future, JLab12 will add very precise DIS data at large $x$, which will allow one
to challenge ideas like helicity retention, predicting that $\Delta f(x)/f(x)\rightarrow 1$ as $x\rightarrow 1$.

Most of the remaining open questions concerning
helicity PDFs are related to their behavior at small $x$
and can be only addressed at a future, high-energy polarized electron-proton collider.
At an EIC, the gluon polarization can be determined precisely from studies of DIS scaling violations \cite{Boer:2011fh}.
A full flavor decomposition down to about $x\simeq 10^{-4}$, including $\Delta s(x)$ and $\Delta \bar{s}(x)$, should be possible by studying the semi-inclusive production of pions and
kaons. If necessary, unpolarized hadron multiplicites will help to constrain FFs better. 
An EIC also has the unique opportunity to access
polarized electroweak structure functions via charged and neutral current DIS measurements.
These novel probes constrain various different combinations of polarized quark PDFs.


\vspace*{0.1cm}
The research of M.S.\ is supported by the U.S.\ Department of Energy under Contract No. 
DE-AC02-98CH10886.

\bibliographystyle{aipproc}   


\end{document}